# Making sound vortices by metasurfaces

Liping Ye,[1] Chunyin Qiu,[1a)] Jiuyang Lu,[1] Kun Tang,[1] Han Jia,[3] Manzhu Ke,[1] Shasha Peng, and Zhengyou Liu[1,2]

[1]Key Laboratory of Artificial Micro- and Nano-structures of Ministry of Education and School of Physics and Technology, Wuhan University, Wuhan 430072, China

[2]Institute for Advanced Studies, Wuhan University, Wuhan 430072, China

[3]State Key Laboratory of Acoustics and Key Laboratory of Noise and Vibration Research, Institute of Acoustics, Chinese Academy of Sciences, Beijing 100190, China

**Abstract**

Based on the Huygens-Fresnel principle, a metasurface structure is designed to generate a sound vortex beam in airborne environment. The metasurface is constructed by a thin planar plate perforated with a circular array of deep subwavelength resonators with desired phase and amplitude responses. The metasurface approach in making sound vortices is validated well by full-wave simulations and experimental measurements. Potential applications of such artificial spiral beams can be anticipated, as exemplified experimentally by the torque effect exerting on an absorbing disk.

[a)]Author to whom correspondence should be addressed. Email:cyqiu@whu.edu.cn

Comparing with optical lenses that are widely used to control light propagation, acoustic lenses have received less attention, especially for audible airborne sound closely related to daily life. The dilemma stems mostly from two critical factors that limit the performance of the acoustic lenses: extreme impedance mismatch and large thickness. The issues are partly relaxed by the recent development of artificial structures,[1-14] such as sonic crystals and acoustic metamaterials, which could provide many acoustic properties unavailable in nature.[1-9] By employing these artificial materials, numerous fascinating phenomena have been observed, for example, negative refractions,[10,11] subwavelength imagings[12,13] and cloakings.[14] Recently, the two-dimensional counterparts of the acoustic metamaterials, i.e., acoustic metasurfaces,[15-20] inspired directly from the pioneering works on optical metasurfaces,[21-25] are attracting great attention due to their capabilities in controlling sound wavefronts. In general, the acoustic metasurfaces are produced by assembling arrays of deep subwavelength building blocks with carefully designed phase and amplitude responses. Flexible wavefront manipulations, e.g., anomalous reflections,[15,16] refractions[17-19] and focusings[15,16,20] have been realized theoretically and experimentally by such flat structures with subwavelength thicknesses. The advantage in thickness enables the metasurface a good candidate to design acoustic components.

Acoustic vortex beams, proposed early by Nye and Berry[26], are attracting extensive interest in recent years.[27-32] Such sound beams, featured by a screwing phase dislocation around the propagation axis, have been anticipated with exciting applications such as acoustic alignment,[27] wave computations,[28] and acoustic spanners for noncontact rotational manipulation.[30,31] Traditionally, the spiral-like sound beams are generated by transducer (or loudspeaker) arrays with controllable phases.[27-31,33-39] Recently, some passive techniques for making sound vortices have also been developed based on the spiral gratings[40,41] and sonic crystals.[42] Here we propose a simple metasurface design, endowed with merits of compact and low cost, to convert a uniform sound beam into the spiral shape wavefront. The metasurface is built by a planar plate decorated with several deep subwavelength resonators engineered with prescribed amplitude and phase responses. The design route has been well confirmed in both full-wave simulations and experimental measurements. Furthermore, the mechanical torque effect exerting on an absorbing disk has been demonstrated preliminarily for such a spiral beam, as an evidence of transferring orbital angular momenta from sound to matter.

It is straightforward to prove that the radiation field of a ring-like line source with constant amplitude and phase gradient $\partial_\theta \phi = n$ can be viewed as an $n$-order vortex beam. The field continuity in the radial direction and the rotational periodicity in the azimuthal direction lead to a spiral-like field pattern with phase singularity in the center. According to the Huygens-Fresnel principle, such an acoustic source could be discretized into a circular array of $m$ point sources as long as the distance among them is in deep subwavelength, associated with a phase difference $2n\pi/m$ with respect to their neighbors. Without losing generality, as shown in Fig. 1(a), in this work a first order vortex beam ($n=1$) is constructed from four point sources ($m=4$) with phases $l\pi/2$ ($l=0, 1, 2, 3$), evenly distributed on a ring of radius $R=0.2\lambda$ (with $\lambda$ being the wavelength in air) in the $z=0$ plane. As shown in Fig. 1(b), the temporal pressure profile and phase pattern in the plane $z=0.2\lambda$ superposed from the four point sources exhibit an excellent spiral field associated with well-defined phase singularity in the vortex core. The spiral field spreads rapidly as $z$ grows, as shown in Fig. 1(c) for $z=2\lambda$. Below we focus on the plane $z=0.2\lambda$, which facilitates the experimental demonstration of the spiral sound field within a finite area.

The above point sources with desired amplitude and phase responses can be mimicked by four subwavelength building blocks, in which the phases emitted must cover a full span of $2\pi$. Different from the early metasurface design to observe anomalous reflections[15,16] and refractions,[17-19] where the wide phase coverage is accomplished by elongating the propagation distance of sound, here the big phase responses are resorted to local resonances. As depicted in the inset of Fig. 2(a), each resonator is made of two large cylindrical cavities (of diameter $D$) connected by three narrow necks (of diameters $d$) drilled through an epoxy plate of height $h$.

Such a unit structure can be viewed as a pair of coupled Helmholtz resonators that enables two resonances in subwavelength region, each resonance leading to a fast growing of phase $\pi$. Assume that the designed operation frequency of the metasurface is 1510 Hz (associated with $\lambda \approx 22.4$ cm) and the plate thickness $h = 9.2$ cm ($\sim 0.39\lambda$). For simplicity, only the heights of the two big cavities $h_1$ and $h_2$ are adjustable and the rest of geometry parameters are fixed as follows: $t_1 = d = 0.60$ cm ($\sim 0.03\lambda$) and $D = 1.84$ cm ($\sim 0.08\lambda$). The length of the neck between the two big cavities can evaluated from $t_2 = h - 2t_1 - h_1 - h_2$ automatically. Since the diameter of the exit $d$ is in deep subwavelength, the sound wave emitting from the resonator can be safely regarded as a point-like source. Throughout the paper, all full-wave simulations are performed by the commercial finite-element solver (COMSOL Multiphysics), where the solid (epoxy) is modeled as acoustically rigid considering the great impedance mismatch with respect to air.

To search the resonators with sound responses prescribed as above, a plane wave is normally incident upon the planar plate perforated with a *single* resonator, in which the rotational symmetry can be applied to reduce the computational time. From the far-field sound response, it is easy to extract the amplitude and phase shift at the exit. (Note that the zero point of phase is fixed at the entrance of the resonator.) As an example, in the inset of Fig. 2(b) we present a set of phase and amplitude spectra for a typical configuration. It is observed that the phase accumulation grows rapidly near the two resonances and covers a wide range of values. Finally, the optimized geometric parameters for the four resonators are: $h_1 = 0.77$ cm, 0.77 cm, 1.76 cm, and 1.76 cm; and $h_2 = 0.77$ cm, 0.86 cm, 1.76 cm and 1.86 cm. In Fig. 2(b), we present the pressure amplitudes and phase responses for the sound emitted from the four resonators (see solid circles). It demonstrates that, as required in the theoretical model, the four discrete phase shifts cover the entire $2\pi$ range and increase with a step of $\pi/2$ among the nearest neighbors, together with a nearly constant transmitted amplitudes (achieved by intentionally choosing configurations near resonances, which is favored to enhance the transmission efficiency of the metasurface.) To further confirm our design, we integrate the four resonators together as in Fig. 1(a) (with $R = 4.7$ cm) and perform a full-wave simulation by incident a plane wave (at 1.51 kHz) normally upon the sample. In Figs. 2(c), we present the instant pressure profile and the corresponding phase pattern extracted for a plane 4.0 cm away from the output interface of the metasurface. It is observed that the full-wave simulation shows clearly the existence of the vortex wavefront with phase singularity in the core, in good agreement with the analytic prediction [see Figs. 1(b)]. The imperfection stems mostly from the coupling among the resonators in real practical design. [We have re-plotted the Figs. 1(b) and 1(c) by using the four point sources with phase and amplitude responses extracted from the full-wave simulations (i.e., the solid circles in Fig. 2(b). Without data presented here, negligible difference is observed comparing with Figs. 1(b) and 1(c).]

Below we present the experimental validation for the predicted spiral field. The experimental setup is depicted in Fig. 3. The cylindrical sample containing four resonators (conveniently prepared by 3D powder printing) is glued with two large parallel Plexiglass plates. The whole system mimics the proposed metasurface, i.e., a single plate (of thickness 9.2cm) perforating with four resonators. It is closely packed with the measurement tube. In experiments, the sound signal is generated directly from a rigid tube of diameter 10 cm. The frequency cutoff for the fundamental waveguiding mode is 1.99 kHz. Therefore, for the frequency concerned here, the incident wave carries uniform amplitude and phase in the cross section, as desired in our design. (Note that the reflection from the sample is unavoidable, but it doesn't affect the wave shape of the incident wave.) The pressure distribution above the whole system can be scanned (with a step size 1.0 cm) by a probe microphone (B&K Type 4187) of diameter 0.7 cm, together with another identical microphone for phase reference. The sound signals launched and received are analyzed by a multi-analyzer system (B&K Type 3560B), from which both of the wave amplitude and phase can be extracted.

At first, we check the sound response for a single resonator by blocking the other ones. The results are presented in Fig. 2(b) for comparison. It is observed that the experimentally measured pressure amplitudes (red open circles) and accumulative phases (blue open circles) for the four resonators agree well with the full-wave simulations. The deviation could be partly from the error in sample fabrication, and partly from the inevitable absorption (due to the viscous effect) that is not considered in the simulation. Then the capability of making a sound vortex is verified with all resonators opened simultaneously. As shown in Fig. 4(a), the sound field scanned above the metasurface displays a well-shaped spiral field at the preset frequency 1.51 kHz, consistent with the numerical prediction in Fig. 2(c) again. A test of the frequency response of the wavefront is necessary for the design route based on subwavelength resonators. In Fig. 4(b) we provide the data for 1.47 kHz. Despite of the field distortion displayed in the instant pressure field, the phase pattern manifests an obvious singularity around the field center, an important feature of the screwing field. This is consistent with the fact that the resonance employed here is not too narrow band [see Fig. 2(b)].

It is well-known that an acoustic vortex beam carries orbital angular momentum.[43,44] When interacting with an absorbing object, the spiral field can transfer orbital angular momentum to the object and hence exert a mechanical torque on it, as confirmed by several pioneering experiments.[31,34,43,45] Potential applications have been anticipated for the acoustically-induced torque effect, such as for noncontact rotational manipulations on particles.[30,31] Here we experimentally demonstrate the capability of rotating objects by the sound vortex sent from the metasurface. As shown in Fig. 5(a), an absorbing foam disk (of radius 5.0 cm and thickness 0.8 cm) is hung 1.0 cm above the structured metasurface by a fine thread. When the sound of 1.51 Hz is launched onto the metasurface, a vortex beam is generated and the foam disk is rotated immediately (see the video in the supplementary material). Finally, the disk is balanced with the torque produced by the twisted thread with more or less rigidity, associated with an angle distinct from the initial one, as shown in Fig. 5(b).

In summary, a metasurface structure has been designed to generate a first-order sound vortex. Good agreement is demonstrated between the sound profiles obtained from full-wave simulations and experimental measurements. Besides, the capability of rotating an absorbing object has been demonstrated experimentally by the vortex beam generated here. The metasurface approach can be conveniently extended to design a vortex of higher order by using more resonant units. It worth pointing out that, the design route proposed here is much simpler than that used for realizing optical vortices by a metasurface,[21] in which a two-dimensional array of resonators are required. Potential applications of such compact metasurface device can be anticipated, such as for the purpose of trapping and rotating objects positioned in the sound field.

**Acknowledgements**

This work is supported by the National Basic Research Program of China (Grant No. 2015CB755500); National Natural Science Foundation of China (Grant Nos. 11374233, 11504275, 11574233 and 11534013).

**Figures and Figure Captions:**

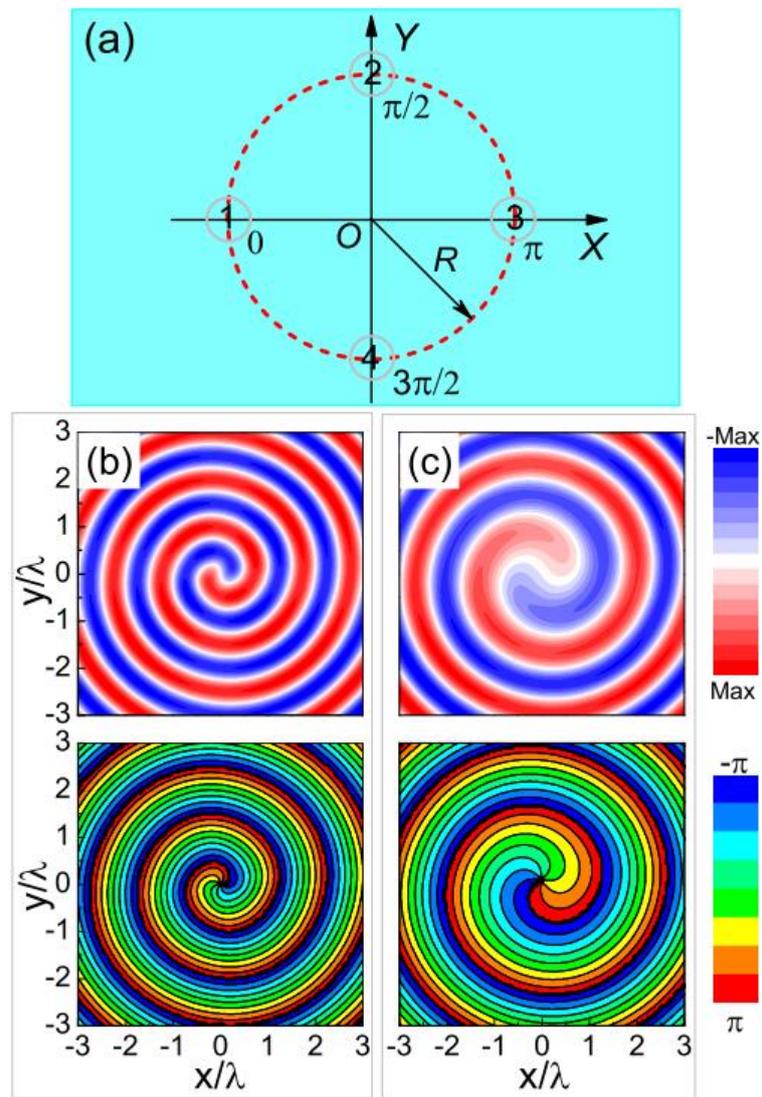

FIG. 1. Vortex design based on the Huygens-Fresnel principle. (a) Spatial distribution of four point sources with constant amplitudes and gradiently varied phases. (b) The instant pressure profile (upper panel) and the corresponding phase pattern (bottom panel) superposed from the four point sources, evaluated for the plane $z = 0.2\lambda$. (c) The same as (b), but for the plane $z = 2\lambda$.

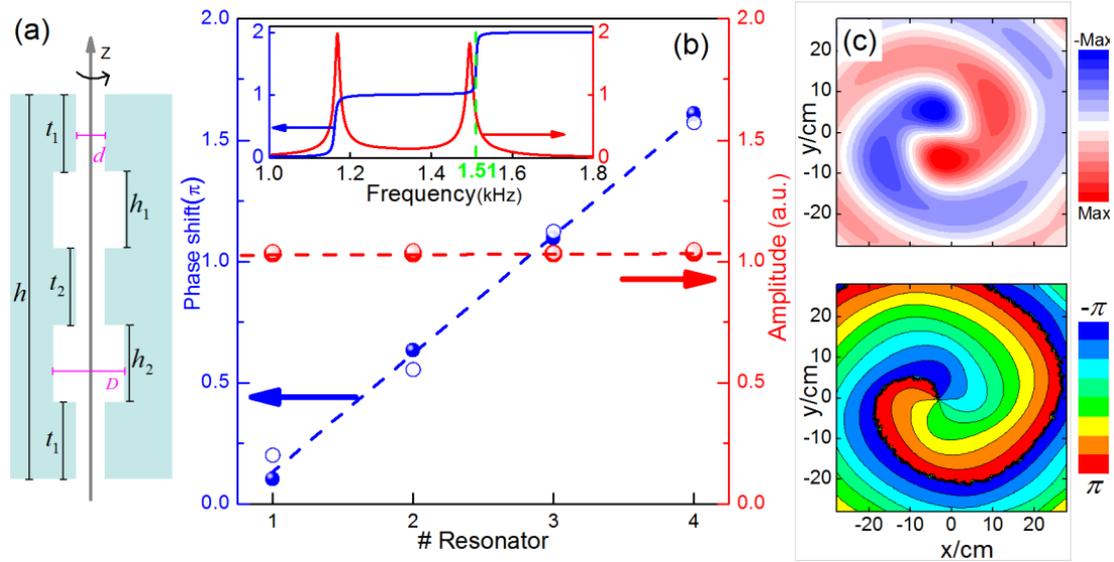

FIG. 2. (a) Cross-sectional view of a resonator (rotated with respect to $z$ axis), consisting of two large circular cavities connected by three narrow necks drilled through an epoxy plate, where only the heights of the two big cavities $h_1$ and $h_2$ are adjustable parameters. (b) Numerically optimized phase (blue solid circles) and amplitude responses (red solid circles) for the four resonators, comparing with the corresponding experiment data (blue and red open circles). Inset: The phase shift (blue) and normalized amplitude (red) spectra for the fourth resonator. (c) The instant pressure profile (upper panel) and phase pattern (bottom panel) simulated by a plane wave incident normally upon the metasurface assembling with four resonators.

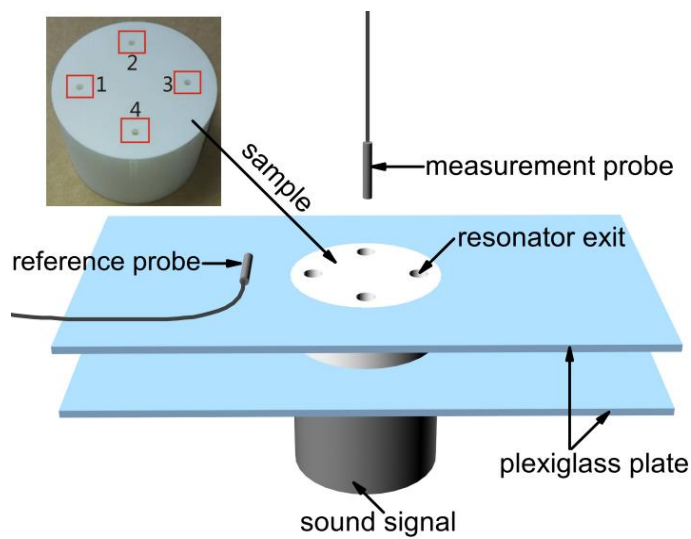

FIG. 3. Experimental setup for generating a sound vortex. The exits of the other resonators will be shielded if the sound response of a single resonator is measured.

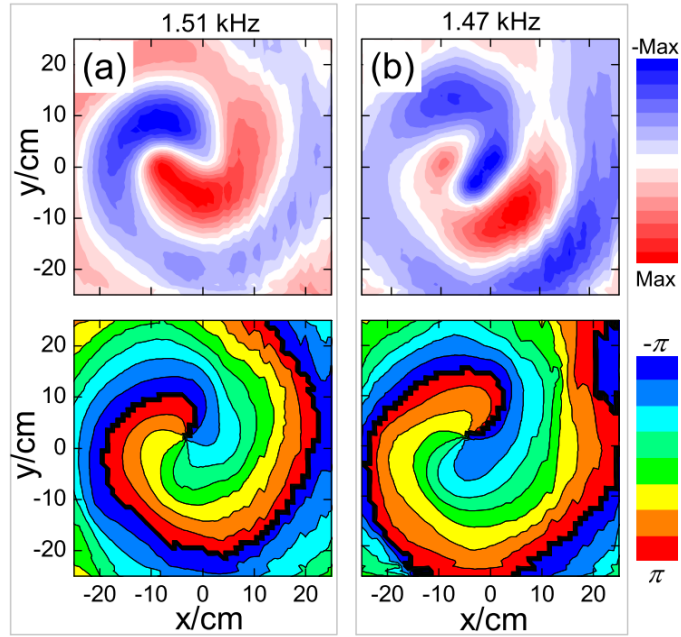

FIG. 4. Instant pressure profiles (top panel) and phase patterns (bottom panel) scanned experimentally on the $z = 4.0$ cm plane, with (a) for 1.51 kHz and (b) for 1.47 kHz.

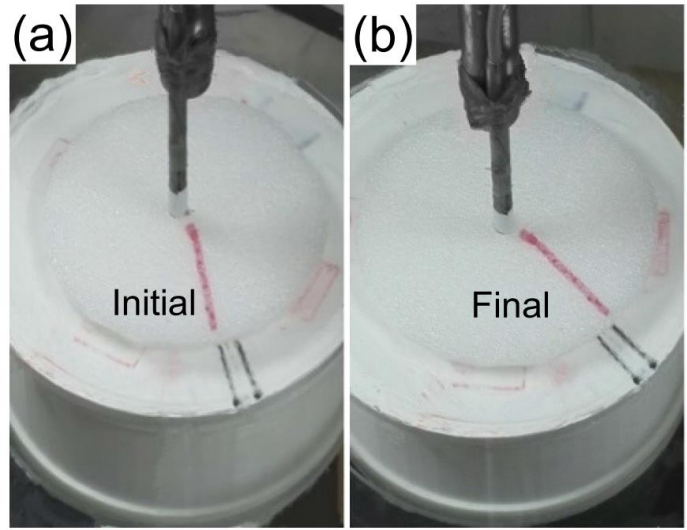

FIG. 5. The torque effect exerted by the vortex beam on an absorbing foam disk that is suspended by a fine thread (not captured well by this top view), where (a) and (b) represent the initial and final states, respectively. Guiding lines are marked for a better observation of the disk rotation.